\documentclass[preprint,aps,nofootinbib]{revtex4}
\usepackage{amsfonts,amsmath,amssymb,amsthm}
\usepackage{latexsym}
\usepackage{bbm,bm}
\usepackage{graphicx}

\newcommand{\bra}[1]{\langle #1 \lvert}
\newcommand{\beq}{\begin{equation}}
\newcommand{\eeq}{\end{equation}}
\newcommand{\beqs}{\begin{eqnarray}}
\newcommand{\eeqs}{\end{eqnarray}}

\begin{document}

\title{Generalized uncertainty principle and Point Interaction}

\author{DaeKil Park$^{1,2}$\footnote{dkpark@kyungnam.ac.kr} and Eylee Jung$^1$,}

\affiliation{$^1$Department of Electronic Engineering, Kyungnam University, Changwon
                 631-701, Korea    \\
             $^2$Department of Physics, Kyungnam University, Changwon
                  631-701, Korea    
                      }

\begin{abstract}
The non-relativistic quantum mechanics with a generalized uncertainty principle (GUP) is examined when the potential is one-dimensional $\delta-$function. 
It is shown that unlike usual quantum mechanics, the Schr\"{o}dinger and Feynman's path-integral approaches are inequivalent at the first order of GUP parameter. 

\end{abstract}

\maketitle

For a lone time it has been believed in the theories of quantum gravity\cite{townsend76,amati89,garay95} that there may exist a minimal observable distance at the Planck scale. 
The modern status of quantum gravity was reviewed in Ref. \cite{rovelli98,carlip01}. The physical motivation for the existence of minimal distance (and/or momentum) is 
due to the conjecture that gravity may disturb the spacetime structure significantly at the Planck scale.
It is also hoped that the introduction of minimal uncertainty may cure the nonrenormalizable character of quantum gravity.

The existence of minimal observable distance and momentum modifies the Heisenberg uncertainty principle (HUP) to the generalized uncertainty principle (GUP).
Thus, the commutation relation between position operator and its conjugate momentum operator should be modified\cite{kempf93,kempf94}. 
Although there are several expressions, we use in this paper the simplest form of the GUP proposed in Ref.\cite{kempf94}
\begin{eqnarray}
\label{gup-1}
&& \left[ \hat{Q}_i, \hat{P}_j \right] = i \hbar \left( \delta_{ij} + \alpha \delta_{ij} \hat{P}^2 + 2 \alpha \hat{P}_i \hat{P}_j  \right)   
                                                                                                                                                                                           \\    \nonumber
&& \hspace{1.0cm} \left[ \hat{Q}_i, \hat{Q}_j \right] = \left[\hat{P}_i, \hat{P}_j \right] = 0
\end{eqnarray}
where the GUP parameter $\alpha$ has a dimension $(\mbox{momentum})^{-2}$. Of course, we return to the HUP if $\alpha = 0$.
Eq. (\ref{gup-1}) is solved up to the first order in $\alpha$ by defining the position and momentum operators as 
\begin{equation}
\label{gup-2}
\hat{P}_i = \hat{p}_i \left(1 + \alpha \hat{p}^2 \right),   \hspace{1.0cm} \hat{Q}_i = \hat{q}_i,
\end{equation}
where $\hat{p}_i$ and $\hat{q}_i$ obey the usual commutation relations of HUP:
\begin{equation}
\label{usual-1}
\left[ \hat{q}_i, \hat{p}_j \right] = i \hbar \delta_{ij}, \hspace{1.0cm} \left[\hat{q}_i, \hat{q}_j \right] = \left[ \hat{p}_i, \hat{p}_j \right] = 0.
\end{equation}
Then, $\hat{P}_i$ and $\hat{Q}_i$ satisfy 
\begin{equation}
\label{gup-3}
 \left[ \hat{Q}_i, \hat{P}_j \right] = i \hbar \left( \delta_{ij} + \alpha \delta_{ij} \hat{p}^2 + 2 \alpha \hat{p}_i \hat{p}_j  \right).
 \end{equation}

The non-relativistic quantum mechanics with GUP-corrected Hamiltonian was examined by Schr\"{o}dinger approach\cite{kempf94} and 
Feynman's path-integral approach\cite{das2012,gangop2019}. In the usual quantum mechanics the transition amplitude $K[q_f, t_f: q_0, t_0]$ from 
$(t_0, q_0)$ to $(t_f, q_f)$, which is usually called Kernel\cite{feynman}, is calculated by a path-integral
\begin{equation}
\label{path-def}
K[q_f, t_f: q_0, t_0] \equiv \bra{q_f, t_f} q_0, t_0 \rangle = \int_{(t_0, q_0)}^{(t_f, q_f)} {\cal D} q e^{(i / \hbar) S[q]},
\end{equation}
where $S[q]$ is an actional functional and ${\cal D} q$ is sum over all possible paths connecting $(t_0, q_0)$ and $(t_f, q_f)$ in spacetime. The Kernel also can be represented as 
\begin{equation}
\label{path-def-2}
K[q_f, t_f: q_0, t_0] = \sum_{n=1}^{\infty} \phi_n (q_f) \phi_n^* (q_0) e^{-(i / \hbar) E_n (t_f - t_0)},
\end{equation}
where $\phi_n (q)$ and $E_n$ are eigenfunction and eigenvalue of Schr\"{o}dinger equation. As far as we know, there is no rigorous 
mathematical proof that the path-integral in Eq. (\ref{path-def}) always results in the right hand side of Eq. (\ref{path-def-2}).
Since, however, no counterexample is found in the usual quantum mechanics, it is asserted that Schr\"{o}dinger and path-integral 
approaches are equivalent. In this short paper, however, we will show that Schr\"{o}dinger and path-integral 
approaches are inequivalent in the non-relativistic GUP-corrected quantum mechanics when the potential is singular $\delta$-function.

In the quantum mechanics with HUP the Feynman propagator for the one-dimensional $\delta$-function potential problem was exactly derived in Ref. \cite{1d-schulman}. 
If, however, one applies the computational technique used in Ref. \cite{1d-schulman} to the higher-dimensional $\delta$-function potential problems, the propagators become infinity.
This is due to the fact that the Hamiltonian with too singular potential loses its self-adjoint property. From the pure aspect of mathematics this difficulty can be 
overcome by incorporating the self-adjoint extension\cite{capri,solvable} into the quantum mechanics. In this way the Schr\"{o}dinger equation for the higher-dimensional 
$\delta$-function potential problems were solved in Ref. \cite{jackiw}. Subsequently, the Feynman propagators and corresponding energy-dependent Green's functions were explicitly derived\cite{park95}. 
From the aspect of physics this difficulty can be overcome by introducing the renormalization scheme to non-relativistic quantum mechanics. In this way same problems were 
reexamined in Ref. \cite{jackiw,huang}. Using renormalization and self-adjoint extension techniques the $\delta'-$ potential problem was also solved\cite{park96,grosche}. 
The equivalence of these two different methods was discussed in Ref. \cite{park97}.

Now, we start with an one-dimensional free particle with GUP, whose Hamiltonian can be written as 
\begin{equation}
\label{free-1}
\hat{H}_0 = \frac{\hat{P}^2}{2 m} = \frac{\hat{p}^2}{2 m} + \frac{\alpha}{m} \hat{p}^4 + {\cal O} (\alpha^2).
\end{equation}
Thus, the time-independent Schr\"{o}dinger equation can be written in a form
\begin{equation}
\label{schrodinger-1}
\left[ -\frac{\hbar^2}{2 m} \frac{d^2}{dq^2} + \frac{\alpha \hbar^4}{m} \frac{d^4}{dq^4} + {\cal O} (\alpha^2) \right] \phi (q) = {\cal E} \phi (q),
\end{equation}
where ${\cal E}$ is an energy eigenvalue. The solution of Eq. (\ref{schrodinger-1}) is 
\begin{equation}
\label{schrodinger-2}
\phi (q) = \frac{1}{\sqrt{2 \pi}} e^{i k q}    \hspace{1.0cm}   {\cal E} = \frac{\hbar^2 k^2}{2 m} + \frac{\alpha \hbar^4}{m} k^4 + {\cal O} (\alpha^2).
\end{equation}
If one extends Eq. (\ref{path-def-2}) to the continuous variables, the corresponding Kernel (or Feynman propagator) can be derived as 
\begin{equation}
\label{kernel-1}
K_0 [q_f, t_f: q_0, t_0] = \int_{-\infty}^{\infty} d k \phi (q_f) \phi^* (q_0) e^{-\frac{i}{\hbar} {\cal E} T},
\end{equation}
where $T = t_f - t_0$. Within ${\cal O} (\alpha)$ the integral in Eq. (\ref{kernel-1}) can be computed and the final expression is 
\begin{eqnarray}
\label{kernel-2}
&&K_0 [q_f, t_f: q_0, t_0] = \sqrt{\frac{m}{2 \pi i \hbar T}} \left[ 1 + \frac{3 i \alpha \hbar m}{T} - \frac{6 \alpha m^2 (q_f - q_0)^2}{T^2} \right]
                                                                                                                                                                      \\    \nonumber
&& \hspace{3.5cm}  \times \exp \left[ \frac{i m (q_f - q_0)^2}{2 \hbar T} \left\{ 1 - 2 \alpha m^2 \left( \frac{q_f - q_0}{T} \right)^2 \right\} \right].
\end{eqnarray}
This expression exactly coincides with the result of Ref. \cite{das2012,gangop2019}, where the Kernel is derived by direct path-integral. Of course, when 
$\alpha = 0$, Eq. (\ref{kernel-2}) reduces to usual well-known free-particle propagator\cite{feynman,kleinert}. 

By introducing the Euclidean time
$\tau = i T$, it is easy to derive the Brownian or Euclidean propagator in a form
\begin{eqnarray}
\label{brownian-1}
&&G_0 [q_f, q_0: \tau] 
= \sqrt{\frac{m}{2 \pi  \hbar \tau}} \left[ 1 - \frac{3  \alpha \hbar m}{\tau} + \frac{6 \alpha m^2 (q_f - q_0)^2}{\tau^2} \right]                                                                                 
\\    \nonumber
&& \hspace{3.5cm}  \times \exp \left[ -\frac{ m (q_f - q_0)^2}{2 \hbar \tau} \left\{ 1 + 2 \alpha m^2 \left( \frac{q_f - q_0}{\tau} \right)^2 \right\} \right].
\end{eqnarray}
Then, the energy-dependent Green's function is defined as a Laplace transform of $G_0 [q_f, q_0: \tau]$:
\begin{equation}
\label{laplace-1}
\hat{G}_0 [q_f, q_0: \epsilon] \equiv \int_0^{\infty} G_0 [q_f, q_0: \tau] e^{-\epsilon \tau} d \tau,
\end{equation}
where $\epsilon = E / \hbar$ and $E$ is an energy parameter. Using an integral formula\cite{table-1}
\begin{equation}
\label{integral-1}
\int_0^{\infty} x^{\nu -1} e^{-\frac{\beta}{x} - \gamma x} d x = 2 \left(\frac{\beta}{\gamma} \right)^{\nu / 2} K_{\nu} \left(2 \sqrt{\beta \gamma} \right),
\end{equation}
where $K_{\nu} (z)$ is a modified Bessel function, one can derive the energy-dependent Green's function up to first order in $\alpha$. The 
final expression is 
\begin{equation}
\label{energy-1}
\hat{G}_0 [q_f, q_0: \epsilon] = \sqrt{\frac{m}{2 \hbar \epsilon}} (1 + 6 \alpha \hbar m \epsilon) 
\exp \left[ - \sqrt{\frac{2 m \epsilon}{\hbar}} \left(1 + 2 \alpha \hbar m \epsilon \right) |q_f - q_0| \right].
\end{equation}

Now, let us consider a problem of $\delta-$function potential, whose Hamiltonian is
\begin{equation}
\label{point-1}
\hat{H}_1 = \hat{H}_0 + v \delta (q).
\end{equation}
Then, the corresponding Schr\"{o}dinger equation is 
\begin{equation}
\label{d-schrodinger-1}
\left[ -\frac{\hbar^2}{2 m} \frac{d^2}{dq^2} + \frac{\alpha \hbar^4}{m} \frac{d^4}{dq^4} + v \delta (q) \right] \phi (q) = E \phi (q).
\end{equation}
Eq. (\ref{d-schrodinger-1}) naturally imposes an boundary condition on $\phi (x)$ at the origin in a form
\begin{equation}
\label{d-boundary-1}
\left[ \phi'(0^+) - \phi' (0^-) \right] - 2 \alpha \hbar^2 \left[ \phi''' (0^+) - \phi''' (0^-) \right] = \frac{2 m v}{\hbar^2} \phi (0).
\end{equation}
If $\alpha = 0$, Eq. (\ref{d-boundary-1}) reduces to usual boundary condition introduced in the Kronig-Penney model.
Also, one can show from Eq. (\ref{d-schrodinger-1}) that if $v < 0$, there is a single bound state. Using $\frac{d}{dx} |x| = \epsilon (x)$, $\frac{d}{d x} \epsilon (x) = 2 \delta (x)$,
and $\delta' (x) f(x) = - \delta (x) f' (x)$, where $\epsilon (x)$ is usual alternating function, one can show that within ${\cal O} (\alpha)$ the normalized bound state $\phi_B (q)$ and bound state energy $B$ are 
\begin{equation}
\label{d-bound-1}
\phi_{B} (q) = \sqrt{a} e^{-a |q|}    \hspace{2.0cm} B = - \frac{m v^2}{2 \hbar^2} - \frac{\alpha m^3 v^4}{\hbar^4},
\end{equation} 
where $a = -m v / \hbar^2 - 2 \alpha m^3 v^3 / \hbar^4 = \sqrt{-2 m B} (1 - 2 \alpha m B) / \hbar$. It is straightforward to show that 
$\phi_B (q)$ satisfies the boundary condition (\ref{d-boundary-1}). 

Now, let us check whether the Kernel or energy-dependent Green's function for $\hat{H}_1$ yields the same results or not. 
Let the Brownian propagator for $\hat{H}_1$ be $\hat{G}[q_f, q_0: \tau]$. Then,  $\hat{G}[q_f, q_0: \tau]$ satisfies an integral equation\cite{feynman,schulman}
\begin{eqnarray}
\label{schwinger}
\Delta G[q_f, q_0: \tau]&=& -\frac{v}{\hbar} \int_0^{\tau} ds \int_{-\infty}^{\infty} dq G_0 [q_f, q: \tau - s] \delta(q) G[q, q_0: s]
                                                                                                                                                             \\     \nonumber
&=&  -\frac{v}{\hbar} \int_0^{\tau} ds G_0 [q_f, 0: \tau - s] G[0, q_0:s],
\end{eqnarray}
where $\Delta G[q_f, q_0: \tau] = G[q_f, q_0: \tau] - G_0[q_f, q_0: \tau]$. Taking a Laplace transform in Eq. (\ref{schwinger}), the energy-dependent Green's function 
$\hat{G}[q_f, q_0: \epsilon]$ for $\hat{H}_1$ satisfies 
\begin{equation}
\label{energy-h1-1}
\Delta \hat{G} [q_f, q_0: \epsilon] = - \frac{v}{\hbar} \hat{G}_0 [q_f, 0:\epsilon] \hat{G} [0, q_0: \epsilon],
\end{equation}
where $\Delta \hat{G} [q_f, q_0: \epsilon] = \hat{G}[q_f, q_0: \epsilon] - \hat{G}_0 [q_f, q_0: \epsilon]$. Solving Eq. (\ref{energy-h1-1}), one can derive
\begin{equation}
\label{energy-h1-2}
\Delta \hat{G} [q_f, q_0: \epsilon] = -\frac{v}{\hbar} \frac{\hat{G}_0[q_f, 0: \epsilon] \hat{G}_0[0, q_0: \epsilon]}{1 + \frac{v}{\hbar} \hat{G}_0 [0, 0: \epsilon]}.
\end{equation}
Inserting Eq. (\ref{energy-1}) into Eq. (\ref{energy-h1-2}), it is possible to derive $\Delta \hat{G} [q_f, q_0: \epsilon]$ in a form
\begin{equation}
\label{energy-h1-3}
\Delta \hat{G} [q_f, q_0: \epsilon] = - \frac{\frac{m}{2 \hbar \epsilon} (1 + 6 \alpha \hbar m \epsilon)^2}
                                                                   {\frac{\hbar}{v} + \sqrt{\frac{m}{2 \hbar \epsilon}} (1 + 6 \alpha \hbar m \epsilon)}
\exp \left[ - \sqrt{\frac{2 m \epsilon}{\hbar}} (1 + 2 \alpha \hbar m \epsilon) (|q_f| + |q_0|) \right].
\end{equation}
Since Eq. (\ref{energy-h1-3}) is valid only up to first order of $\alpha$, $\Delta \hat{G}$ can be expressed as 
\begin{eqnarray}
\label{revise-1}
&&\Delta \hat{G} [q_f, q_0: \epsilon] =  -\frac{v}{\hbar} \sqrt{\frac{m}{2 \hbar \epsilon}} \exp \left[ - \sqrt{\frac{2 m \epsilon}{\hbar}} (|q_f| + |q_0|) \right]                                 \\     \nonumber
&&\hspace{1.0cm} \times \left[ \frac{1}{\frac{v}{\hbar} + \sqrt{\frac{2 \hbar \epsilon}{m}}} \left\{ 1 + 12 \alpha \hbar m \epsilon - 2 \alpha \hbar m \epsilon \sqrt{\frac{2 m \epsilon}{\hbar}} (|q_f| + |q_0| ) \right\}
- \frac{6 \alpha v m \epsilon}{\left(\frac{v}{\hbar} + \sqrt{\frac{2 \hbar \epsilon}{m}}\right)^2}   + {\cal O} (\alpha^2)     \right].
\end{eqnarray}

Now, we derive the Brownian propagator for $\hat{H}_1$ by taking an inverse Laplace transform in Eq. (\ref{revise-1}). Using the following formulas of the inverse Laplace transform\cite{integral}
\begin{eqnarray}
\label{inverse-f}
&& {\cal L}^{-1} \left[ \frac{e^{-a \sqrt{\epsilon}}}{\sqrt{\epsilon} + b} \right] (\tau) = \frac{1}{\sqrt{\pi \tau}} e^{-\frac{a^2}{4 \tau}} - b e^{a b + b^2 \tau} \mbox{erfc} \left(\frac{a}{2 \sqrt{\tau}} + b \sqrt{\tau} \right)
                                                                                                                                                                                                                                                     \\   \nonumber
&& {\cal L}^{-1} \left[ \frac{\epsilon^{-1/2}e^{-a \sqrt{\epsilon}}}{\sqrt{\epsilon} + b} \right] (\tau) =  e^{a b + b^2 \tau} \mbox{erfc} \left(\frac{a}{2 \sqrt{\tau}} + b \sqrt{\tau} \right)
                                                                                                                                                                                                                                                     \\   \nonumber
&& {\cal L}^{-1} \left[ \frac{\epsilon^{1/2} e^{-a \sqrt{\epsilon}}}{\sqrt{\epsilon} + b} \right] (\tau) = \frac{a - 2 b \tau}{2 \sqrt{\pi \tau^3}} e^{-\frac{a^2}{4 \tau}} + b^2 e^{a b + b^2 \tau} \mbox{erfc} \left(\frac{a}{2 \sqrt{\tau}} + b \sqrt{\tau} \right)                                                                                                                                                                                                                                                     
                                                                                                                                                                                                                                                      \\   \nonumber
&& {\cal L}^{-1} \left[ \frac{\epsilon e^{-a \sqrt{\epsilon}}}{\sqrt{\epsilon} + b} \right] (\tau) = \frac{4 b^2 \tau^2 - 2 (a b + 1) \tau + a^2}{4 \sqrt{\pi \tau^5}} e^{-\frac{a^2}{4 \tau}} - b^3 e^{a b + b^2 \tau} \mbox{erfc} \left(\frac{a}{2 \sqrt{\tau}} + b \sqrt{\tau} \right)
                                                                                                                                                                                                                                                      \\    \nonumber
&& {\cal L}^{-1} \left[ \frac{e^{-a \sqrt{\epsilon}}}{(\sqrt{\epsilon} + b)^2} \right] (\tau) = -2 b \sqrt{\frac{\tau}{\pi}} e^{-\frac{a^2}{4 \tau}} + (2 b^2 \tau + a b + 1) e^{a b + b^2 \tau} \mbox{erfc} \left(\frac{a}{2 \sqrt{\tau}} + b \sqrt{\tau} \right)
\end{eqnarray}
where $\mbox{erfc}(z)$ is an error function defined as 
\begin{equation}
\label{error}
\mbox{erfc} (z) = \frac{2}{\sqrt{\pi}} \int_z^{\infty} e^{-t^2} d t,
\end{equation}
then, $\Delta G[q_f, q_0 : \tau] \equiv G[q_f, q_0: \tau] - G_0 [q_f, q_0: \tau]$ can be written in a form
\begin{eqnarray}
\label{brown-delta-1}
&&\Delta G[q_f, q_0 : \tau]                                                                                                     \\      \nonumber
&&= - \frac{m v}{2 \hbar^2} \Bigg[ \exp \left[ \frac{m v}{\hbar^2} \left( |q_f| + |q_0| + \frac{v \tau}{2 \hbar} \right) \right] \mbox{erfc} \left[ \sqrt{\frac{m}{2 \hbar}} \left\{ \frac{|q_f| + |q_0|}{\sqrt{\tau}} + \frac{v}{\hbar} \sqrt{\tau} \right\} \right]
                                                                                                                                                                                                                                                     \\    \nonumber
&&\hspace{3.0cm}\times \left[ 1 + \frac{12 \alpha m^2 v^2}{\hbar^2} + \frac{4 \alpha m^3 v^3 (|q_f| + |q_0|)}{\hbar^4} + \frac{3 \alpha m^3 v^4}{\hbar^5} \tau \right]                         \\    \nonumber
&&\hspace{2.0cm} + \sqrt{\frac{2 \hbar}{ \pi m}} \alpha \exp \left[ - \frac{m (|q_f| + |q_0|)^2}{2 \hbar \tau} \right]                                                                                                   \\    \nonumber
&&\hspace{2.0cm}\times \Big[ - \frac{3 m^3 v^3}{\hbar^4} \tau^{1/2} - \frac{m^2 v}{\hbar} \left( 9 + \frac{m v (|q_f| + |q_0|)}{\hbar^2} \right) \tau^{-1/2}                                                                   \\    \nonumber
&&\hspace{2.0cm}+ m^2 (|q_f| + |q_0|) \left( 7 +  \frac{m v (|q_f| + |q_0|)}{\hbar^2} \right) \tau^{-3/2} 
- \frac{m^3  (|q_f| + |q_0|)^3}{\hbar} \tau^{-5/2}      \Big]     \Bigg].
\end{eqnarray}
The Kernel for $\hat{H}_1$ can be explicitly derived from Eq. (\ref{brown-delta-1}) by changing $\tau \rightarrow i T$.  

If $\alpha = 0$, $\Delta G[q_f, q_0 : \tau]$ reduces to 
\begin{eqnarray}
\label{special-1}
\Delta G[q_f, q_0 : \tau] &=& - \frac{mv}{2 \hbar^2} \exp \left[ \frac{m v}{\hbar^2} \left( |q_f| + |q_0| + \frac{v \tau}{2 \hbar} \right) \right] \mbox{erfc} \left[  \sqrt{\frac{m}{2 \hbar}} \left\{ \frac{|q_f| + |q_0|}{\sqrt{\tau}} + \frac{v}{\hbar} \sqrt{\tau} \right\} \right]
                                                                                                                                                                                                                                                                                           \nonumber      \\
&=& - \frac{m v}{\hbar^2} \int_0^{\infty} d z e^{-\frac{m v}{\hbar^2} z} G_F \left[ |q_f|, -|q_0| - |z|; \tau \right],
\end{eqnarray}
where $G_F [q_f, q_0: \tau]$ is a Brownian propagator for free particle case in usual quantum mechanics, whose explicit form is 
\begin{equation}
\label{usual-1}
G_F [ q_f, q_0: \tau] = \sqrt{\frac{m}{2 \pi \hbar \tau}} e^{- \frac{m (q_f - q_0)^2}{2 \hbar \tau}}.
\end{equation}
Eq. (\ref{special-1}) exactly coincides with the result of Ref. \cite{1d-schulman}. 

In order to explore the equivalence of Schr\"{o}dinger and path-integral approaches we first examine the boundary condition at the origin. Using Eq. (\ref{brown-delta-1}) and 
$\frac{d}{dz} \mbox{erfc}(z) = - \frac{2}{\sqrt{\pi}} e^{-z^2}$, one can show straightforwardly that within ${\cal O} (\alpha)$, $G[q_f, q_0: \tau]$ satisfies 
\begin{eqnarray}
\label{d-boundary-11}
&&\left\{ \frac{\partial}{\partial q_f} G[0^+, q_0: \tau] - \frac{\partial}{\partial q_f} G[0^-, q_0: \tau] \right\} - 
4 \alpha \hbar^2 \left\{ \frac{\partial^3}{\partial q_f^3} G[0^+, q_0: \tau] - \frac{\partial^3}{\partial q_f^3} G[0^-, q_0: \tau] \right\}    \nonumber   \\
&&\hspace{5.0cm}  = \frac{2 m v}{\hbar^2} G[0, q_0: \tau].
\end{eqnarray}
Since this is different from Eq. (\ref{d-boundary-1}) at the first order of $\alpha$, one can say that the Schr\"{o}dinger and path-integral approaches are inequivalent
at the same order of $\alpha$. Of course, two boundary conditions are exactly the same when $\alpha = 0$.  

The difference also arises from the energy-dependent Green's function $\hat{G} [q_f, q_0: \epsilon]$.
Since the pole and residue of the energy-dependent Green's function have an information about the bound state energy and the corresponding bound state, 
Eq. (\ref{energy-h1-3}) implies that if $v < 0$, $\hat{H}_1$ generates a single bound state energy $B'$, whose explicit form is  
\begin{equation}
\label{d-bound-11}
B' \equiv -\hbar \epsilon = - \frac{m v^2}{2 \hbar^2} - \frac{3 \alpha m^3 v^4}{\hbar^4}.
\end{equation}
This is also different\footnote{Since Eq. (\ref{energy-h1-3}) is valid only within ${\cal O} (\alpha)$, one can change the denominator of $\Delta \hat{G} [q_f, q_0: \epsilon]$, and hence its pole. In this case, however, there are 
multiple bound states, which is also different from the result of Schr\"{o}dinger approach.} from second equation of Eq. (\ref{d-bound-1}) at the order of $\alpha$. 
The corresponding normalized bound state is
\begin{equation}
\label{d-bound-12}
\Phi_B (q) = \sqrt{a'} e^{-a' |q|},
\end{equation}
where $a' = -m v/ \hbar^2 - 4 \alpha m^3 v^3 / \hbar^4$. This is also different from bound state derived from the Schr\"{o}dinger equation at the first order of $\alpha$. 
It is straightforward to show that $\Phi_B (q)$ does not obey the boundary condition (\ref{d-boundary-1}) but obeys (\ref{d-boundary-11}).

It is shown that the Schr\"{o}dinger equation and Feynman path-integral do predict different results in the non-relativistic quantum mechanics with GUP when the 
potential is one-dimensional $\delta$-function. This difference may be due to the singular nature of potential. However, one-dimensional $\delta-$function 
potential is well defined in the usual quantum mechanics. Thus, the equivalence of Schr\"{o}dinger and Feynman's path-integral approaches should be 
checked in more detail when the GUP is involved in the non-relativistic quantum mechanics.


\end{document}